\documentclass[5p]{./style/elsarticle/elsarticle}
\usepackage{graphicx}
\usepackage{multirow} 
\usepackage{amssymb}
\usepackage{amsmath}
\usepackage{enumerate}
\usepackage{color}
\usepackage{ulem}
\usepackage[switch]{lineno}
\usepackage{url}
\graphicspath{{../figures/}}
\usepackage{enumitem}

\begin{document}

\begin{frontmatter}

\title{Effect of the diffusion parameters on the observed $\gamma$-ray spectrum of sources and their contribution to the local all-electron spectrum: the EDGE code}


\author[MPIK]{R. L\'opez-Coto}
\ead{rlopez@mpi-hd.mpg.de}
\author[MPIK]{J. Hahn}
\author[UR]{S. BenZvi}
\author[MPIK]{J. Hinton}
\author[UR]{M.U. Nisa}
\author[MPIK]{R. D. Parsons}
\author[IFJ]{F. Salesa Greus}
\author[LANL]{H. Zhou}

\address[MPIK]{Max-Planck-Institut f\"ur Kernphysik, P.O. Box 103980, D 69029 Heidelberg, Germany}
\address[UR]{Department of Physics \& Astronomy, University of Rochester, Rochester, NY, USA}
\address[IFJ]{Instytut Fizyki Jadrowej im Henryka Niewodniczanskiego Polskiej Akademii Nauk, Krakow, Poland}
\address[LANL]{Physics Division, Los Alamos National Laboratory, Los Alamos, NM 87545, USA}

\begin{abstract}
The positron excess measured by PAMELA and AMS can only be explained if there is one or several sources injecting them. Moreover, at the highest energies, it requires the presence of nearby ($\sim$hundreds of parsecs) and middle age (maximum of $\sim$hundreds of kyr) sources. Pulsars, as factories of electrons and positrons, are one of the proposed candidates to explain the origin of this excess. To calculate the contribution of these sources to the electron and positron flux at the Earth, we developed EDGE (Electron Diffusion and Gamma rays to the Earth), a code to treat the propagation of electrons and compute their diffusion from a central source with a flexible injection spectrum. Using this code, we can derive the source's gamma-ray spectrum, spatial extension, the all-electron density in space, the electron and positron flux reaching the Earth and the positron fraction measured at the Earth. We present in this paper the foundations of the code and study how different parameters affect the gamma-ray spectrum of a source and the electron flux measured at the Earth. We also studied the effect of several approximations usually performed in these studies.
\end{abstract}

\begin{keyword}
Gamma-ray astronomy, cosmic ray electrons, positrons, diffusion 
\end{keyword}
\end{frontmatter}


\section{Introduction}
\label{sec:intro}

Cosmic rays (CRs) are high-energy charged particles that strike the atmosphere almost isotropically. They are composed by protons and helium nuclei (99\%), heavier nuclei, electrons (e$^-$), positrons (e$^+$) and other particles in smaller proportions. Currently, there are galactic propagation models that reproduce the e$^\pm$ CR energy spectrum assuming they are secondary products of the sea of CR collisions. 
However, the e$^+$ content in the total e$^\pm$ flux above a few GeV (also known as ``positron excess'') measured by PAMELA and AMS amongst others, can only be explained if there is one or several sources injecting them. Moreover, taking into account standard diffusion and cooling of e$^\pm$, the extension of the e$^\pm$ spectrum up to TeV energies can only be explained if the source is nearby ($\sim$hundreds of parsecs) and middle age ($\sim$hundreds of kyr). Pulsars, as factories of e$^\pm$, are one of the proposed candidates to explain the origin of this excess, however there are also more exotic explanations such as galactic jets \cite{Gupta2014} or dark matter \cite{Ibarra2013}. To measure the maximum energy reached by these e$^\pm$ at the Earth and unveil the origin of the positron excess is one of the most important questions unsolved in astroparticle physics nowadays. Since e$^\pm$ are charged particles, their arrival direction does not point to their origin because they are deflected by magnetic fields. One of the ways to study sources of CRs is to analyse the neutral subproducts of CR collisions such as gamma rays.

We developed EDGE (Electron Diffusion and Gamma rays to the Earth), a code to treat diffusion of electrons that was used in \cite{Geminga_Science} to compute the electron diffusion from a central source, derive its $\gamma$-ray spectrum, profile and the all-electron density in the space. We present in this paper the foundations of the code and study how different parameters affect the $\gamma$-ray spectrum of a source and the electron flux measured at the Earth.


\section{Diffusion of electrons}
\label{sec:diffusion}
If we assume a spherically symmetric case where electrons are diffusing from a central source, the equation that describes this process is:

\begin{equation}
\frac{\partial f}{\partial t} = \frac{D}{r^2}\frac{\partial}{\partial r}r^2\frac{\partial f}{\partial r}  + \frac{\partial}{\partial \gamma}(Pf) + Q
\end{equation}

where $\gamma$=E/m$_e$c$^2$ with E the energy of the particle, m$_e$ the mass of the electron and c the speed of light. $f(r,t,\gamma)$ is the energy distribution of particles at an instant t and distance r from the source, $D(\gamma$) the energy dependent diffusion coefficient, $P(\gamma$) the energy loss rate and $Q(r,t,\gamma$) is proportional to the injection spectrum. The details about how to solve this equation for particular cases can be found in  \cite{Atoyan95}. The Green function for this equation for an arbitrary injection spectrum $\Delta N (\gamma)$ is given by:

\begin{equation}
\label{eq:energy_density}
 f(r,t,\gamma) = \frac{\Delta N (\gamma_t) P(\gamma_t)}{\pi^{3/2}P(\gamma) r_{\textrm{diff}}^3} \exp \left( - \frac{r^2}{r_{\textrm{diff}}^2} \right)
\end{equation}

where $\gamma_t$ corresponds to the initial energy of the particles. The diffusion radius ($r_{\textrm{diff}}$) represents the mean free path of e$^\pm$ of a given energy. It is given by:

\begin{equation}
\label{eq:rdiff}
r_{\textrm{diff}}=2\sqrt{\Delta u}
\end{equation}

and $\Delta u$:

\begin{equation}
\Delta u = -\int^{\gamma}_{\gamma_t}{  D(x) dx / P(x) } 
\label{eq:lambda}
\end{equation}

is the integral over the particle history from an initial energy $\gamma_t$ to an energy $\gamma$. 

\subsection{Injection spectrum}
In the literature, three types of time-dependency on the injection spectrum have widely been used: burst-like, continuous and pulsar-like injection. 

\subsubsection{Burst-like injection}
The burst-like injection is not suitable to describe the Very High Energy (VHE; $E>100$ GeV) gamma-ray spectrum of a source with an age of the order of the sources we are interested. The problem is that the electrons producing VHE gamma-ray emission are very energetic and they cool down very fast. An approximation for the cooling time of a particle when this is dominated by Inverse Compton (IC) and synchrotron losses in the Thomson regime is:

\begin{equation}
\label{eq:cooling}
t_{\rm{cool}} \approx 3 \times 10^5 \left( \frac{E}{\rm{TeV}} \frac{\epsilon}{\rm{eV\ cm}^{-3}} \right)^{-1} \rm{yr}
\end{equation}

where $E$ the energy of the electrons and $\epsilon$ is the energy density of the target photons plus the magnetic field. The electrons producing VHE $\gamma$-ray emission ($\gtrsim$ 1 TeV) have a cooling time smaller than the age of the pulsar, producing a low level of VHE $\gamma$-ray emission that cannot fit the observations.

\subsubsection{Continuous injection}
The continuous injection scenario has the advantage of having an analytical solution given in \cite{Atoyan95}:

\begin{equation}
\label{eq:energy_density_continuous_injection}
 f_{\rm{cont}}(r,t,\gamma) = \frac{Q_0 \gamma^{-\alpha}}{4\pi D(\gamma) r} \rm{erfc} \left( \frac{r}{2\sqrt{D(\gamma) t_\gamma}} \right)
\end{equation}

this scenario is not realistic in the case of the injection of a pulsar either, which has a more complicate dependence with time. On the other hand, it is useful to derive the $\gamma$-ray spectrum of the source at energies $\gtrsim$10 TeV for sources with $t_{\rm{age}}\gtrsim10^5$ yr. The VHE $\gamma$-ray emission is produced by freshly accelerated electrons ($\gtrsim$10 TeV electrons have t$_{\rm{cool}}\lesssim10^4$ years). This approach is on the other hand not suitable to describe the propagation of electrons to the Earth. 

\subsubsection{Pulsar-like injection}
\label{sec:time-dependence}

Pulsars are rotating neutron stars that produce periodic radiation by spinning their powerful magnetic field through space. They loss their rotational energy by emitting a wind of electron and positron pairs that diffuses away when these particles escape outside of the pulsar's magnetosphere. We assume that this emission is isotropic and the wind is composed by the same quantity of electrons and positrons. Let us talk about some properties of pulsars that are important to characterize their emission.

The luminosity of a pulsar is given by \cite{Gaensler06}:
\begin{equation}
\label{eq:luminosity}
L(t) = L_0\left(   1 +  \frac{t}{\tau_0}\right)^{-\frac{n+1}{n-1}}
\end{equation}

where $n$ is the braking index of the pulsar, $L_0$ is the initial luminosity and $\tau_0$ the initial spin-down timescale. We assume that the pulsar behaves as a dipole, therefore $n$=3. The age of the system is given by:

\begin{equation}
\label{eq:age}
t_{\rm{age}} = \frac{P}{(n-1)\dot{P}} \left[  1-\left( \frac{P_0}{P} \right)^{n-1} \right]
\end{equation}

where $P$,$\dot{P}$ and $P_0$ are the period, period derivative and birth period of the pulsar respectively. The characteristic age $\tau_{\rm{c}}$ of a pulsar is estimated from its period and period derivative and is given by:

\begin{equation}
\tau_{\rm{c}} = \frac{P}{2 \dot{P}}
\end{equation}



If we assume that the spectrum of injected electrons is given by a power-law:

\begin{equation}
\label{eq:injection_spectrum}
\frac{dN}{dE}=Q(\gamma,t) = Q_0 \gamma^{-\alpha}
\end{equation}

where $Q_0$ is the initial injection rate and $\alpha$ the injection rate's index. The injection rate is related to the pulsar's luminosity by the equation:

\begin{equation}
L_e(t) = \int^{\gamma \rm{min}}_{{\gamma \rm{max}}}Q(\gamma,t) \gamma m_{\rm{e}} c^2 d\gamma
\end{equation}

with $L_e(t)=\mu L(t)$. $L(t)$ is given by equation \ref{eq:luminosity} and $\mu$ is a constant $<$1 that determines the fraction of the luminosity that is transferred to electrons. The initial injection rate is therefore given by:

\begin{equation}
Q_0 = \left( \int^{\gamma \rm{min}}_{{\gamma \rm{max}}}\gamma^{-\alpha} \gamma m_{\rm{e}} c^2 d\gamma \right)^{-1} \mu L_0\left(   1 +  \frac{t}{\tau_0}\right)^{-2}
\end{equation}

\subsection{Energy loss}
The energy loss rate is given by: 

\begin{equation}
\label{eq:energyloss}
P(\gamma) = -\frac{d \gamma}{dt}
\end{equation}

here we include synchrotron, IC and bremsstrahlung losses.

\subsubsection{Synchrotron:}

The synchrotron losses are given by \cite{Moderski05}:

\begin{equation}
P(\gamma)_{\rm{syn}} = \frac{4 \sigma_T c}{3} \frac{B^2}{8\pi} \gamma^2
\end{equation}

with $B$ the magnetic field experienced by the particles.

\subsubsection{Inverse Compton:} We calculate the IC cross-section in two different regimes, depending on the target photon ($E_{\rm{ph}}$) and electron ($E_e$) energies:
\label{sec:inverse_compton}
\begin{enumerate}

\item if ($E_{\rm{ph}} \gamma_e / (m_e c^2)  < 0.1 $) 

we use the Thomson equation for the IC losses:

\begin{equation}
P(\gamma)_{\rm{IC}} = \frac{4 \sigma_T c}{3} U_{\rm{ph}} \gamma^2
\end{equation}

with $\sigma_T$ the Thomson IC cross-section, $c$ the speed of light and $U_{\rm{ph}}$ the energy density of the target photons.

\item if ($E_{\rm{ph}} \gamma_e / (m_e c^2)  > 0.1 $) 

P($\gamma_{\rm{IC}}$) is calculated using the full Klein-Nishina treatment of the electron-photon cross section \cite{Blumenthal70}. We use equation 2.48 of \cite{Blumenthal70} to calculate the emissivity of the radiation fields and integrate it over the photon and electron spectra. The radiation targets are added as a grey body distribution with mean the temperature of each of the radiation fields.
\end{enumerate}

\subsubsection{Bremsstrahlung:}

The bremsstrahlung energy losses are taken from Eq. 9 of \cite{Haug04} for $E> 500$ keV. The ambient density assumed is $n_{\rm{H}}$=1~cm$^{-3}$
 

\subsection{Diffusion coefficient}
The energy-dependent diffusion coefficient is given by:

\begin{equation}
\label{eq:diff_coefficient}
D(\gamma) = D_{0} \left(   1+ \frac{\gamma}{\gamma^{\star}}  \right)^\delta
\end{equation}

where D$_{0}$ is the diffusion coefficient normalization, $\gamma^{\star}$ is usually taken $\sim$ 6$\times$10$^3$ (corresponding to $E^{\star}$=3 GeV) and $\delta$ the diffusion index, usually between 0.3 and 0.6 \cite{Atoyan95,Yuksel08}.

\subsection{Electron/positron flux at the Earth}
\label{sec:electron_positron_flux}
The electron flux at the Earth for a given source with an age $t_{\rm{age}}$ and situated at a distance $d$ is given by:

\begin{equation}
\label{eq:flux_earth}
J(\gamma) = \frac{c}{4\pi} f(d_{\rm{Earth}},t_{\rm{age}},\gamma)
\end{equation}

where $d_{\rm{Earth}}$ is the distance from the source to the Earth.

\subsection{Positron fraction at the Earth}
\label{sec:electron_positron_fraction}

The positron fraction for a given $\gamma$ will be given by:

\begin{equation}
\label{eq:fraction_earth}
\frac{\rm{e}^+}{\rm{e}^++\rm{e}^-} (\gamma)= \frac{0.5 \times \rm{J(\gamma) + Sec.[e}^+]}{\rm{J(\gamma)+ Sec.[e}^+]+\rm{Prim.[e}^-]+\rm{Sec.[e}^-]}
\end{equation}

where the primary electrons are considered to be injected by astrophysical sources, and secondary electron and positron fluxes are products of CR collisions. For these fluxes, we use the phenomenological curves from the right panel of Figure 5 of \cite{Moskalenko97}.


\section{Electron, positron and $\gamma$-ray fluxes calculations using EDGE}
\label{sec:edge}

The \textbf{E}lectron \textbf{D}iffusion and \textbf{G}amma rays at the \textbf{E}arth (\textbf{EDGE}) is a flexible code that accepts as input parameters different pulsar and environment characteristics and gives as a final output the $\gamma$-ray spectrum produced by the source and the electron and positron flux produced at the Earth by this source.  
The code uses dependencies from the GAMERA package \cite{GAMERA}. We will give an overview of the calculations performed by the code and the results obtained with a set of selected parameters.

\subsection{Default paremeters}
In the following, we will give a brief description of the default parameters used in the code. 

\subsubsection{General parameters}
We will be studying the properties of a known pulsar through its known parameters and assumed values. The pulsar selected is Geminga, a middle-age (characteristic age  $\tau_c=342$ kyr) pulsar located at a distance to the Earth $d_{\rm{Earth}}$=250 pc \cite{manchester}. The period of the pulsar is $P=237$~ms and its period derivative $\dot{P}$=1.10$\time10^{-14}$. The spin-down power of the pulsar is $\dot{E}=3.2\times10^{34}$ erg/s. 

\subsubsection{Injection spectrum}

Since we would like to derive a $\gamma$-ray spectrum together with the all-electron spectrum at the Earth, we will use the pulsar-like injection mechanism described in Section \ref{sec:time-dependence}. 
The minimum and maximum energy of the simulated electron spectrum are chosen $E_{\rm{min}}$=1 GeV and $E_{\rm{max}}$=500 TeV, respectively. The injection spectrum index assumed is $\alpha$=2.2. The fraction of spin-down power transformed into $\gamma$-ray emission assumed is $\mu$=0.5.

\subsubsection{Energy losses}
The target photon fields used are:
\begin{itemize}

\item CMB: $\epsilon_{\rm{CMB}}=0.26\ $eV/cm$^3$, T=2.7 K
\item Infrarred: $\epsilon_{\rm{IR}}=0.3\ $eV/cm$^3$, T=20 K
\item Optical: $\epsilon_{\rm{Opt}}=0.3\ $eV/cm$^3$, T=5000 K
\end{itemize}
The default value for the magnetic field is $B=3 \mu$G.

\subsubsection{Diffusion parameters}
We select $\delta$=0.33 driven by \cite{Kolmogorov1941} and in agreement with recent measurements \cite{Aguilar2016_BCratio}.
The normalization $D_{0}$=4$\times10^{27}$ cm$^{-2}$s$^{-1}$ as in \citep{Yuksel08}.

\subsection{Energy density in space}
\begin{figure}
\begin{center}
\includegraphics[width=0.48\textwidth]{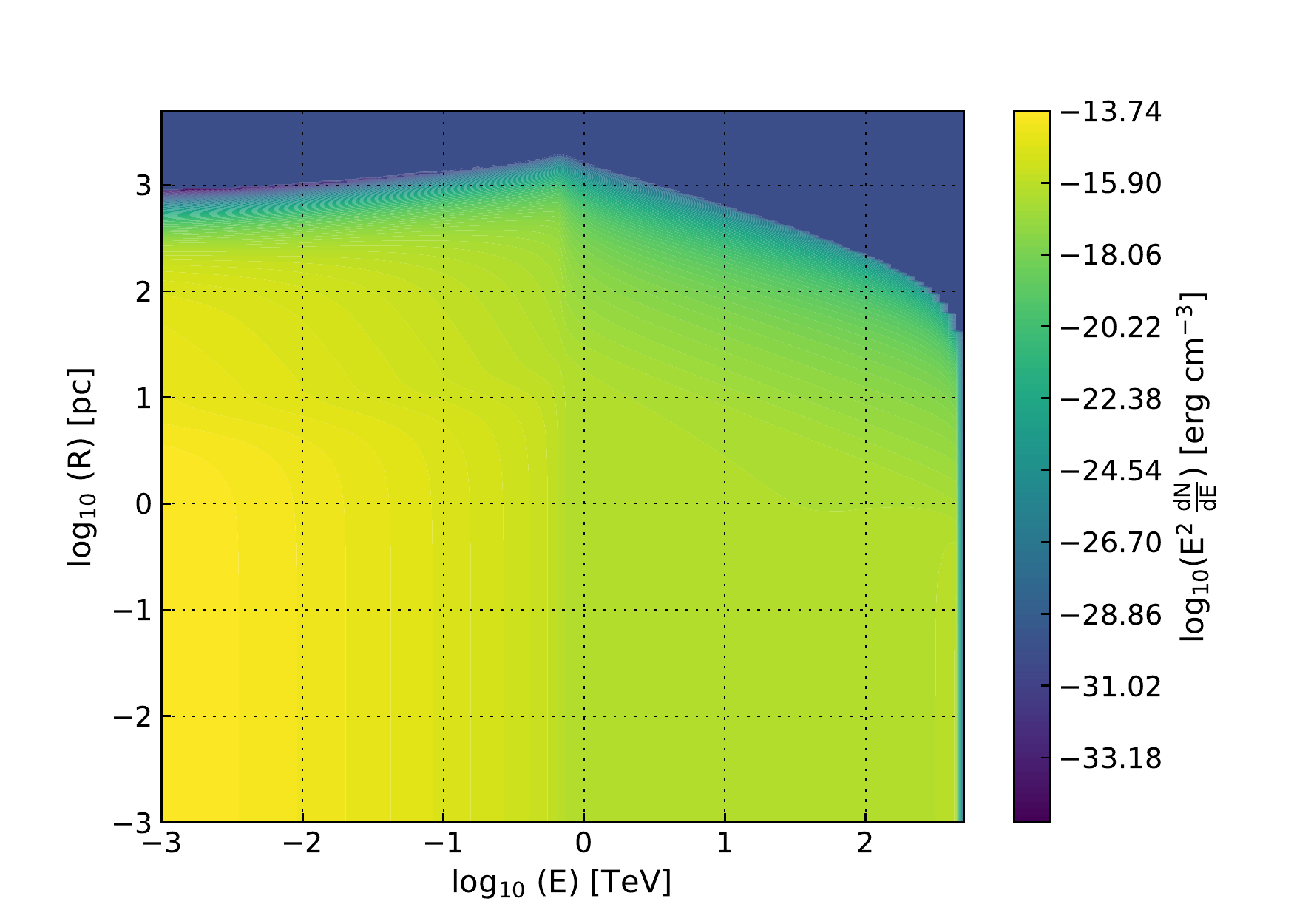}
\caption{Energy density of the e$^\pm$ population emitted by the central pulsar as a function of the distance from the pulsar $R$ and their current energy $E$.}
\label{fig:energy_density}
\end{center}
\end{figure}

The procedure to get the electron and positron flux at the Earth starts by evaluating the energy density of the electrons produced by the central pulsar (Eq. \ref{eq:energy_density}) for a time $t=t_{\rm{age}}$ in every point of the space and for the full range of energies. If we numerically solve Eq. \ref{eq:lambda} and \ref{eq:energyloss} and substitute them in Eq. \ref{eq:energy_density}, we obtain a look-up table with the energy density of e$^\pm$ for different energies at different distances from the pulsar as it is represented on Figure \ref{fig:energy_density}. The Figure's shape reflects the expectation we have from Eq. \ref{eq:rdiff}. If we consider cooling in the Thomson regime, and the approximation that the $r_{\textrm{diff}}\approx 2\sqrt{D(\gamma)t_{\rm{d}}}$ we can substitute Eq. \ref{eq:cooling} to calculate the diffusion radius. Depending on the energy, the diffusion radius will have the following dependencies:

\begin{enumerate}[label=\roman*)]

\item if $t_{\rm{cool}} (E) < t_{\rm{age}}$ (corresponding to $E\gtrsim1$ TeV), the system is cooling-limited, $t_{\rm{d}}=t_{\rm{cool}}$ and the diffusion radius $r_{\rm{diff}}\propto E^{(\delta-1)/2}$ increases with decreasing energy.

\item if $t_{\rm{cool}} (E) > t_{\rm{age}}$ (corresponding to $E\lesssim1$ TeV), the system is age-limited, $t_{\rm{d}}=t_{\rm{age}}$ and the diffusion radius is $r_{\rm{diff}}\propto E^{\delta/2}$ increases with increasing energy.

\end{enumerate}

\subsection{Electron spectra}
\label{sec:electron_spectra}

\begin{figure}
\begin{center}
\includegraphics[width=0.48\textwidth]{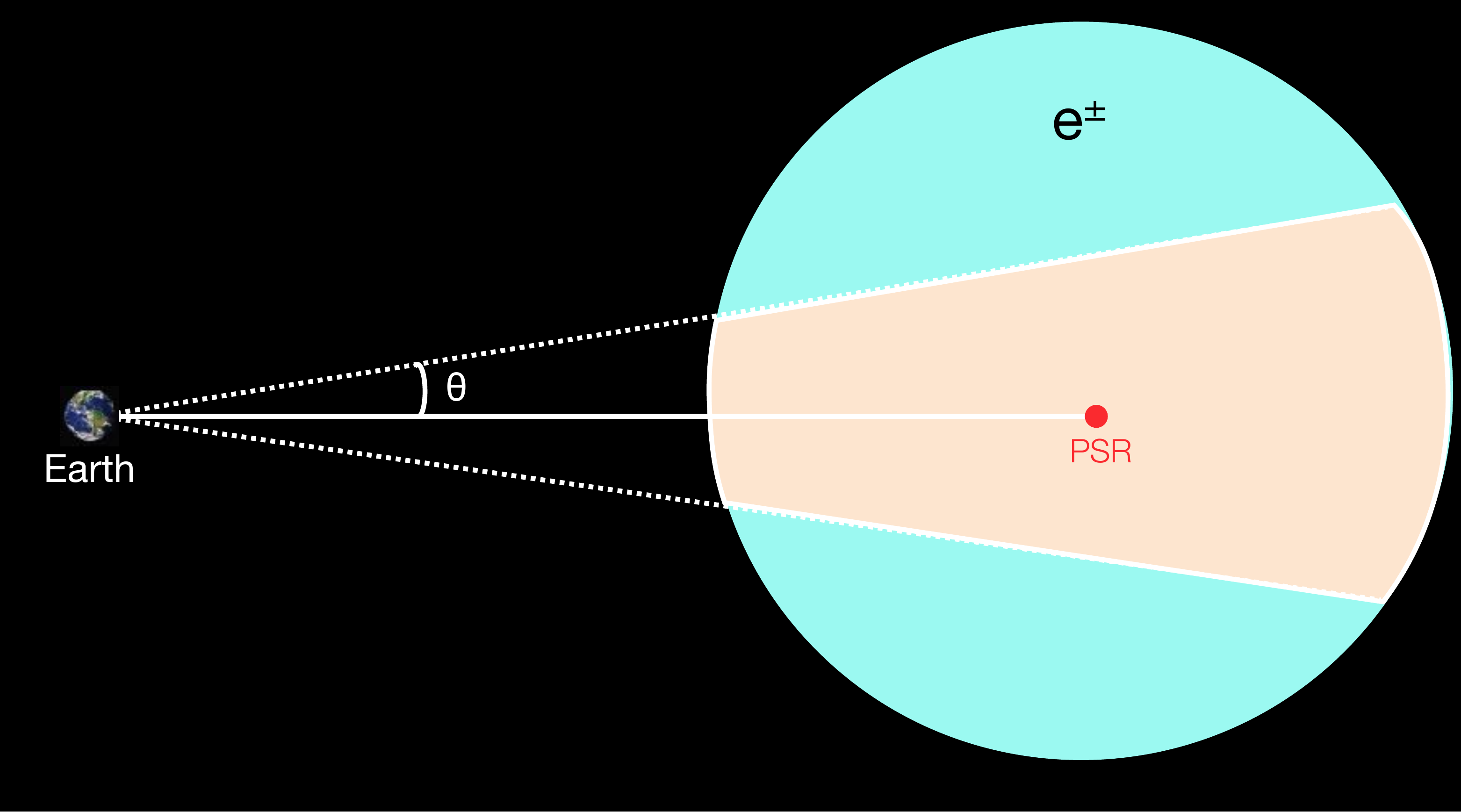}
\caption{Line of sight integral from the Earth. The red point at the center corresponds to the position of the pulsar. For a given distribution of electrons denoted by the cyan sphere, and a size of the source $\theta$, the electrons producing the $\gamma$-ray emission we detect at the Earth are the ones included inside the brown region.}
\label{fig:lineofsight}
\end{center}
\end{figure}

Once we have the energy density of electrons in all points of the space, we compute the electrons that are inside the volume given by the line of sight from the earth to the source. For a size $\theta$ of the source, we need to integrate all the electrons inside the brown region shown on Figure \ref{fig:lineofsight}. 

\begin{figure}
\begin{center}
\includegraphics[width=0.48\textwidth]{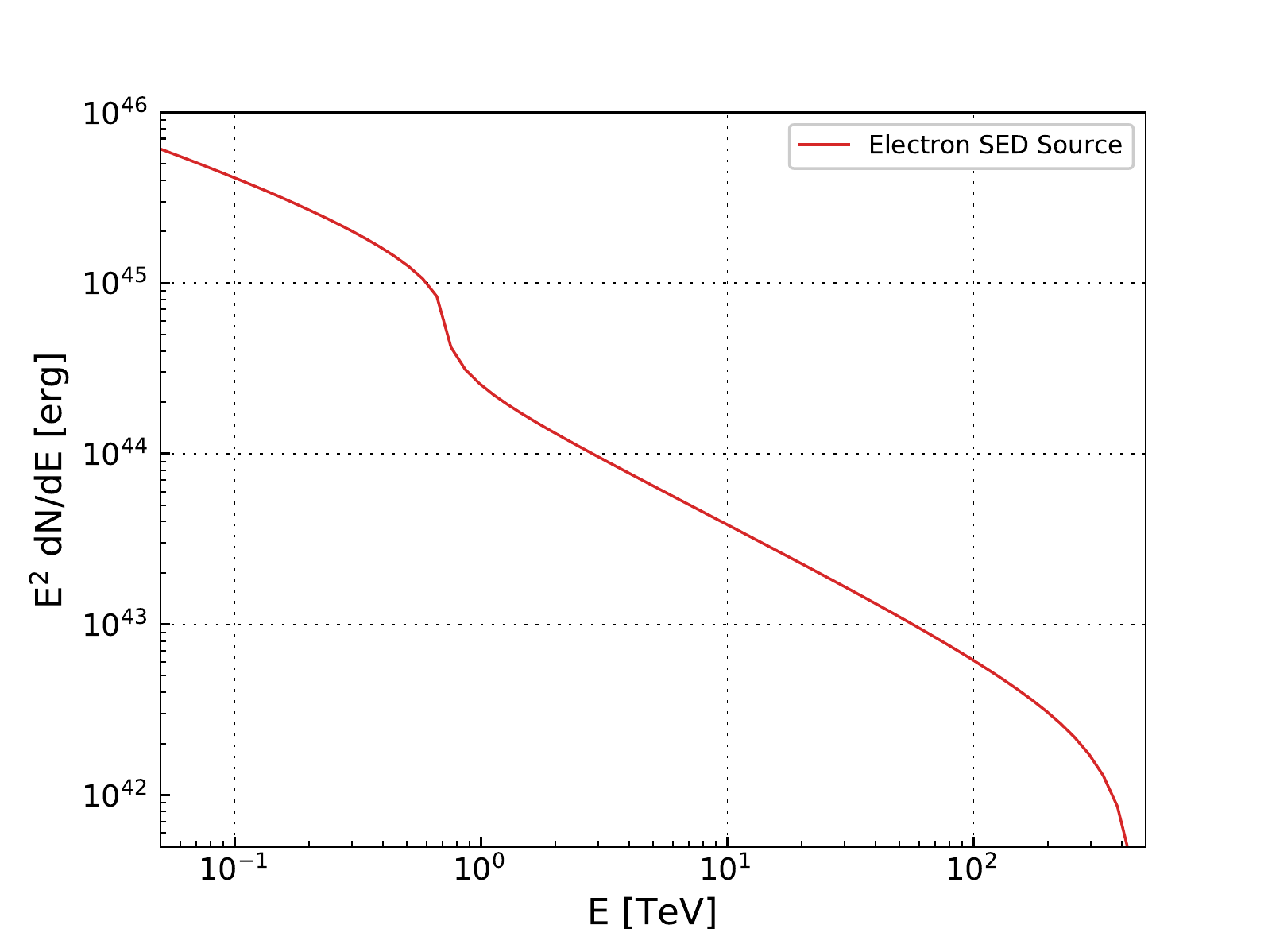}
\caption{Spectral energy distribution of electrons contained inside the volume of the source. }
\label{fig:electron_spectra}
\end{center}
\end{figure}

The selected size for the source is $\theta=5^\circ$. We compute the differential energy of the electrons contained inside the volume delimited by the cone shown in two dimensions on Figure \ref{fig:lineofsight}. To do it, we integrate over a sphere centered at the pulsar position with its boundaries limited by the size of the cone. The result of this integration is the differential energy spectrum of all the electrons that are producing the $\gamma$-ray emission. The spectral energy distribution of these electrons is shown in Figure \ref{fig:electron_spectra}. The kink present at $E\sim1$ TeV represents the transition from the energy where electrons are not cooled to that where they are cooled. The difference in spectral indices between these two regions has the same origin. 


\subsection{Gamma-ray spectra}
\begin{figure}
\begin{center}
\includegraphics[width=0.48\textwidth]{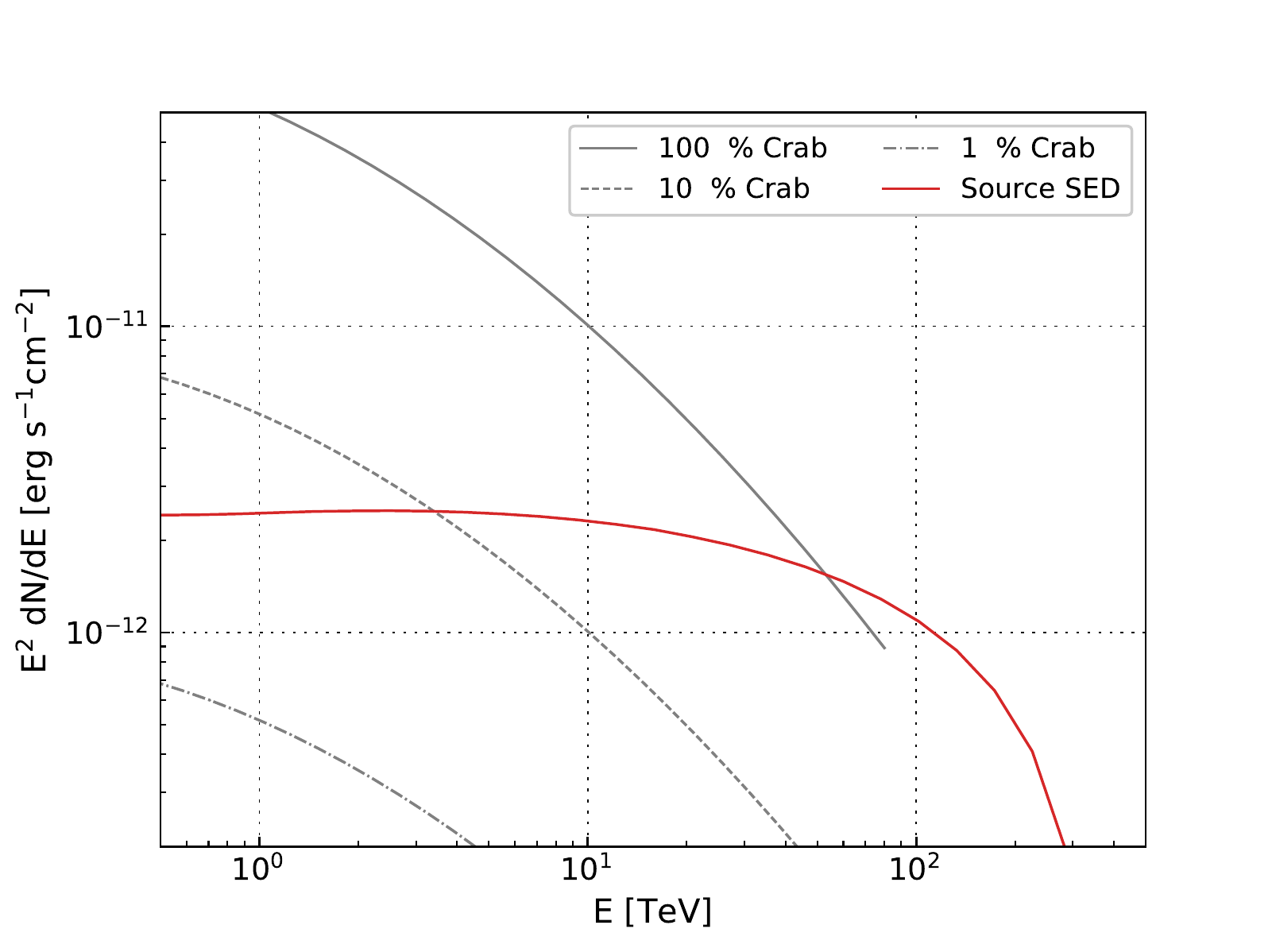}
\caption{Gamma-ray spectra for a source of $\theta=5^\circ$. Gray lines showing the Crab flux are also included.}
\label{fig:gamma_spectra}
\end{center}
\end{figure}

The gamma-ray spectrum at TeV energies is produced by IC up-scattering of ambient photons, mainly CMB at multi-TeV energies. As mentioned in Section \ref{sec:inverse_compton}, we use Eq. 2.48 from \cite{Blumenthal70} to calculate the IC emissivity of the electrons shown in Figure \ref{fig:electron_spectra}. The results for the VHE $\gamma$-ray spectrum are shown in Figure \ref{fig:gamma_spectra}.

\subsection{Gamma-ray spatial profiles}

\begin{figure}
\begin{center}
\includegraphics[width=0.48\textwidth]{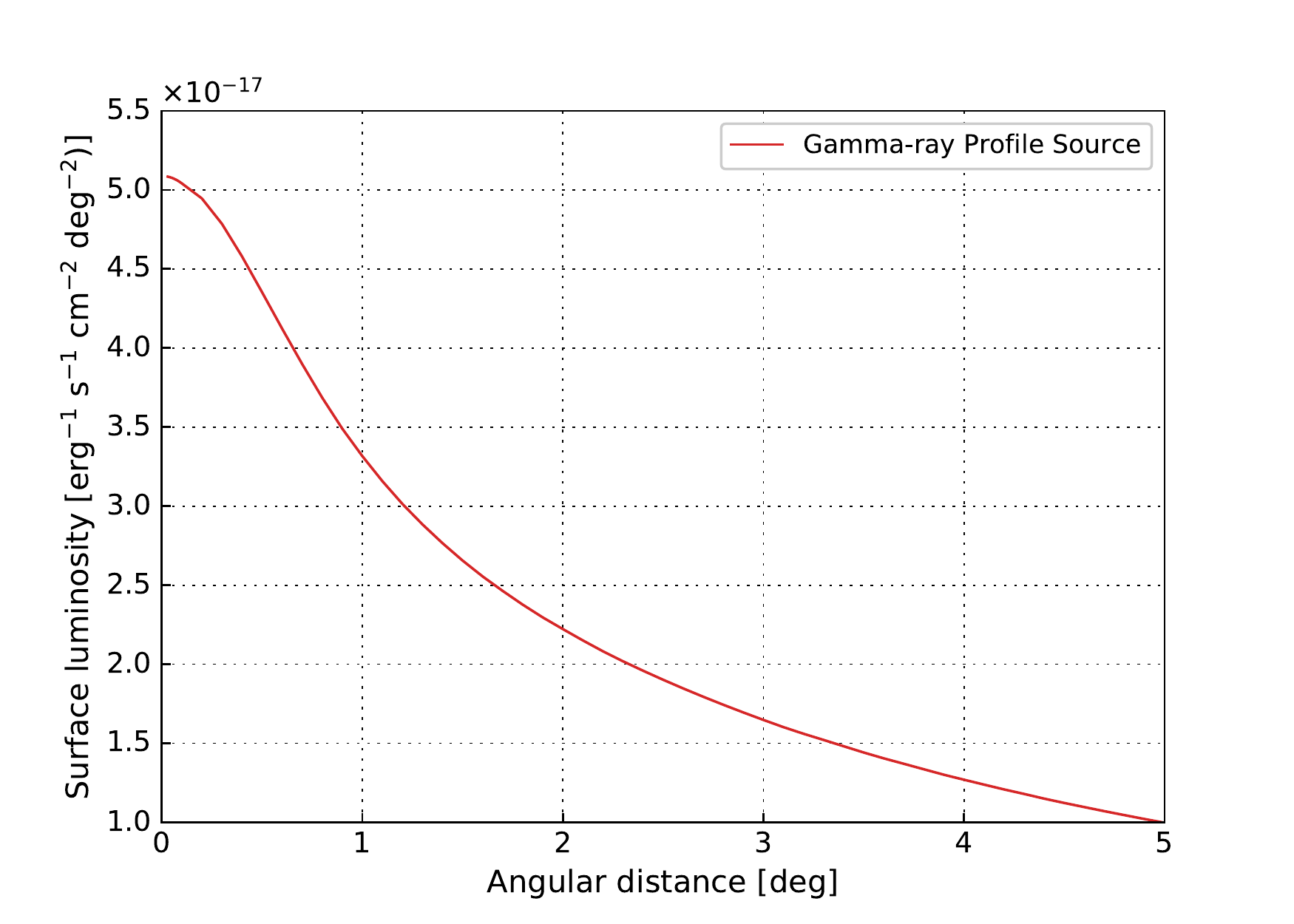}
\caption{Differential flux per solid angle at 20 TeV.}
\label{fig:gamma_spectra_profiles}
\end{center}
\end{figure}

We can also calculated the $\gamma$-ray angular profiles at different distances from the source. We show the differential flux per solid angle at 20 TeV for different angular distances from the pulsar on Figure \ref{fig:gamma_spectra_profiles}. 

\subsection{Modeling $e^\pm$ Propagation to Earth}

\begin{figure}
\begin{center}
\includegraphics[width=0.48\textwidth]{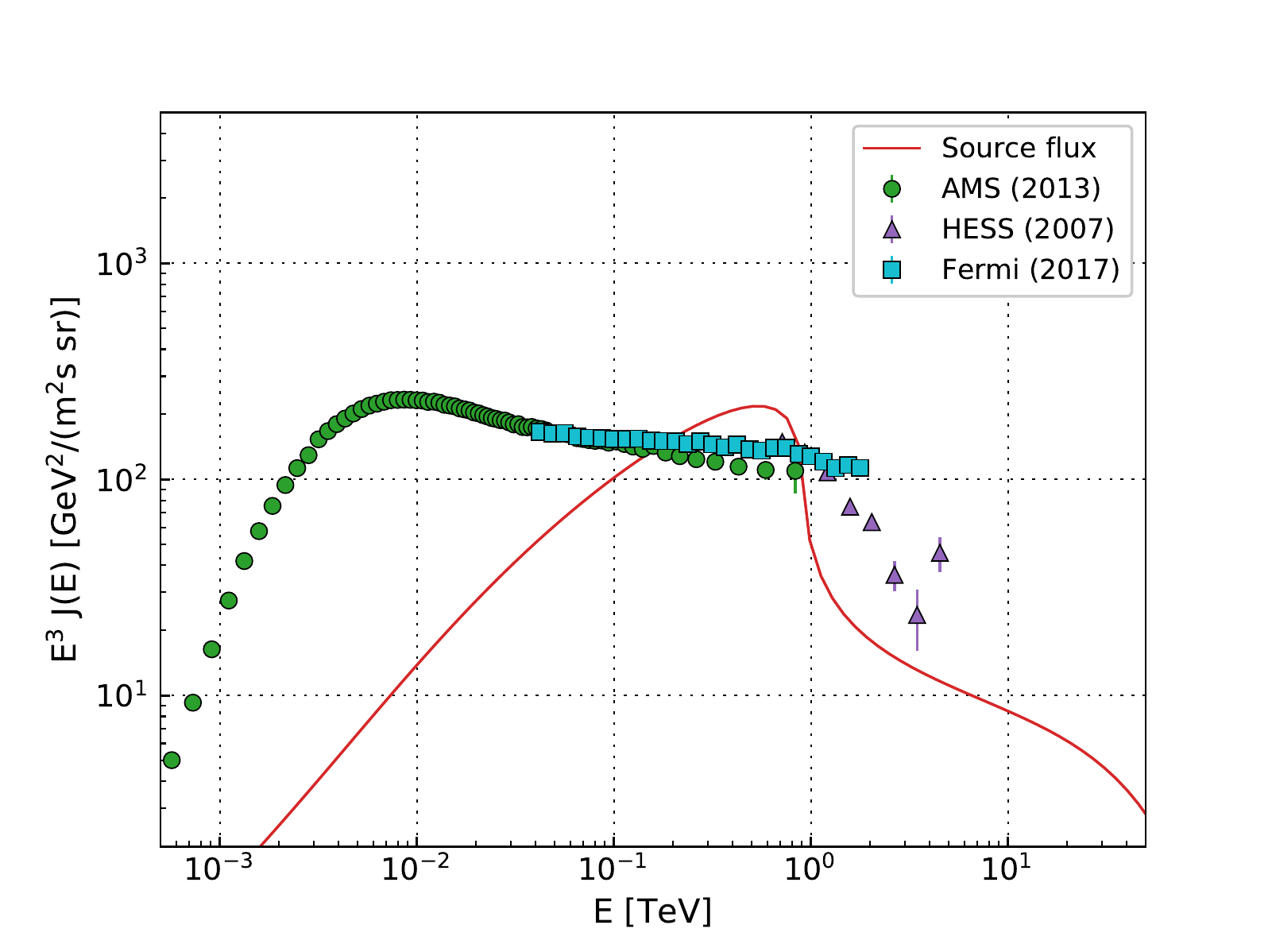}
\includegraphics[width=0.48\textwidth]{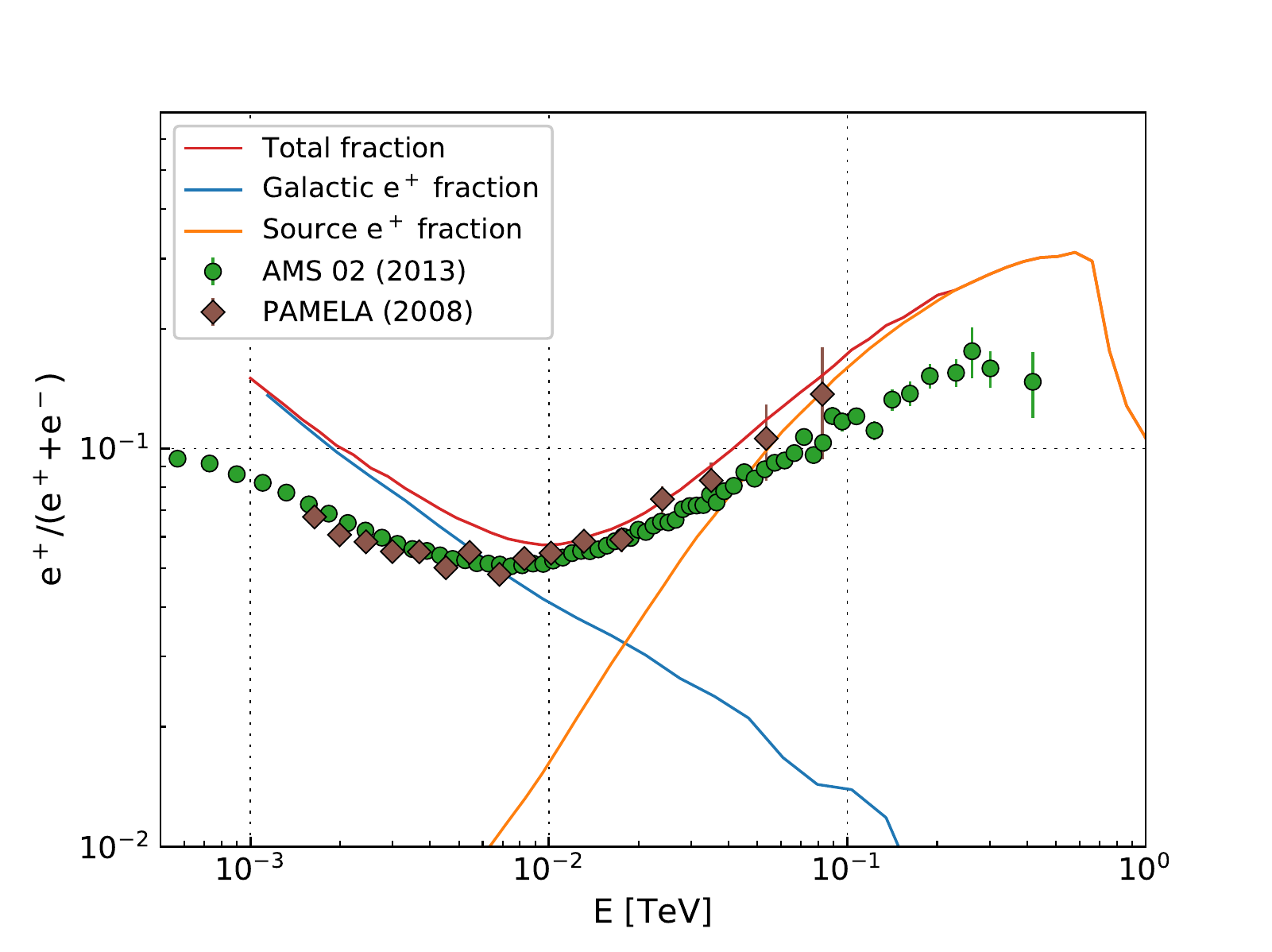}
\caption{Electron/positron flux at the Earth (top panel) and positron fraction (bottom panel).}
\label{fig:flux_fraction_Earth}
\end{center}
\end{figure}

To compute the electron and positron flux provided by a given source at the Earth, we use Eq. \ref{eq:flux_earth}. For the fraction of positrons that this source contributes at the Earth, we use Eq. \ref{eq:fraction_earth}. The results are shown on Figure \ref{fig:flux_fraction_Earth}. Note that for the default parameters selected for this example, the all-electron spectrum at the Earth produced by the pulsar overshoots the measured one for energies of $\sim$hundreds of GeV. This indicates that the selected parameters for the diffusion do not correspond to the real conditions of the Interstellar Medium (ISM).

\section{Discussion}

We will discuss the effect of different approximations and also how all the different parameters of the model affect the local all-electron flux and the VHE $\gamma$-ray flux of the source.

\subsection{Effect of particle suppression}

\begin{figure}
\begin{center}
\includegraphics[width=0.48\textwidth]{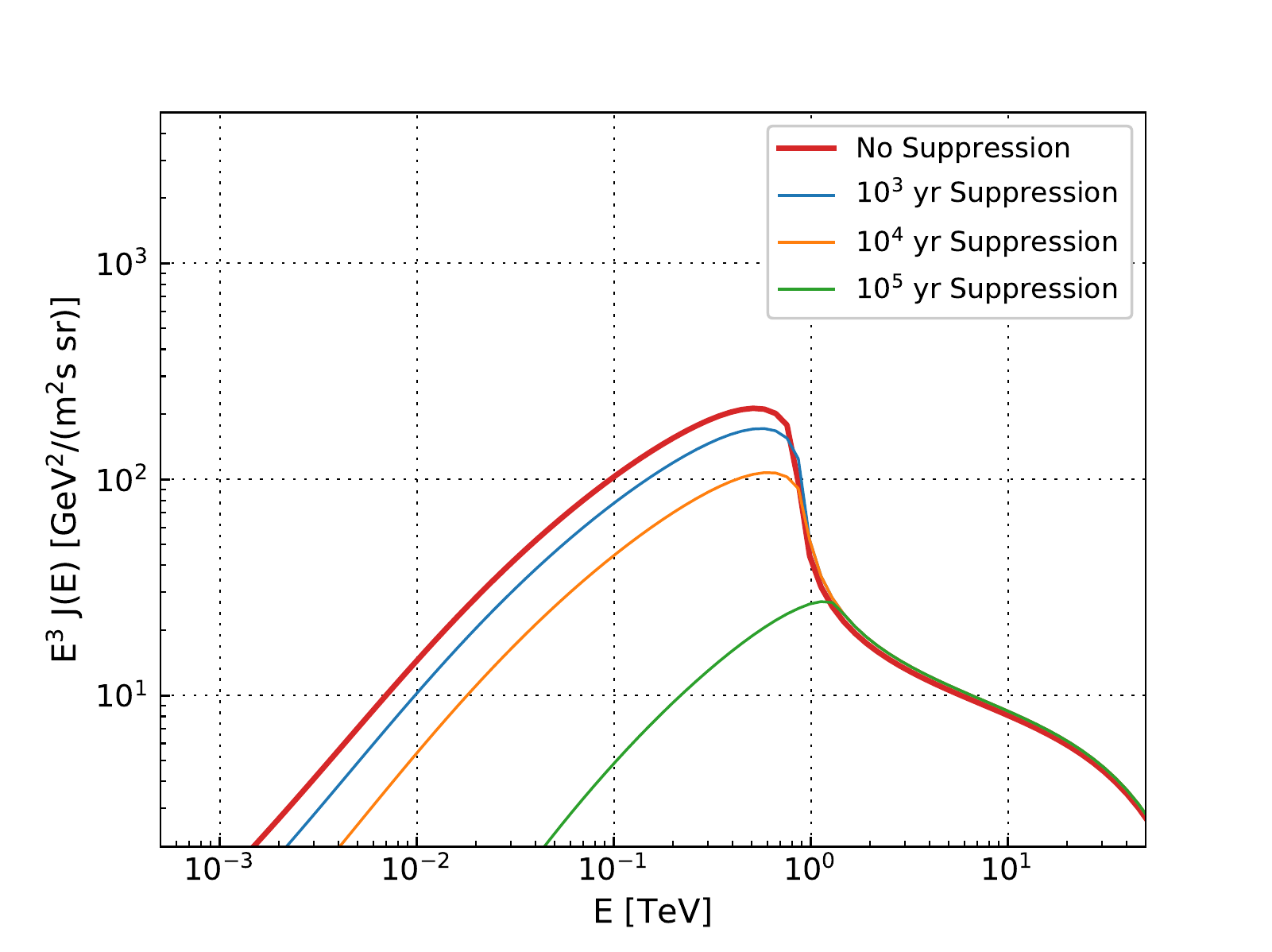}
\caption{All-electron (e$^-$+e$^{+}$) flux at the Earth applying different suppression times for the particles injected.}
\label{fig:confinement_effects_gamma_electrons}
\end{center}
\end{figure}

In our model, we consider that particles diffuse into the ISM as soon as they are injected by the pulsar. In reality, these particles spend some time confined within the PWN before they are allowed to escape. To evaluate this effect, we calculated the e$^\pm$ at the Earth for different confinement times and we show the results in Figure \ref{fig:confinement_effects_gamma_electrons}. For simplification, we consider that the confined particles lose all their energy in the confinement time and are therefore suppressed. Only particles injected afterwards contribute to the e$^\pm$ flux at the Earth. The VHE $\gamma$-ray spectrum produced by the source is not affected since it is the product of very high energy electrons that are injected relatively recently, compared to the suppression times considered.

The e$^\pm$ flux at the Earth is reduced, specially for confinement times of the order of the age of the pulsar. We can conclude that for confinement times larger than $\sim10^4$ yr, where the bulk of the pulsar power is emitted, the all-electron flux at the Earth is only slightly affected, while the VHE $\gamma$-ray flux remains unaffected.

\subsection{Effect of approximations in the electron cooling}

We evaluated the effect of using approximations on the energy losses as the ones described in equation 14 of \cite{Atoyan95}:

\begin{equation}
P(\gamma)=p_0 + p_1 \gamma + p_2 \gamma^2
\end{equation}

where $p_0=6\times10^{-13} n$ s$^{-1}$ corresponds to the ionization losses,  $p_1=10^{-15} n$ s$^{-1}$ to the bremsstrahlung and $p_2=5.2\times10^{-20}$ $w_0$ s$^{-1}$ to the IC and synchrotron energy losses, with $n$ defined as the particle density in cm$^{-3}$ and $w_0$ the addition of the different target photon and magnetic field energy densities in eV cm$^{-3}$.

\begin{figure}
\begin{center}
\includegraphics[width=0.48\textwidth]{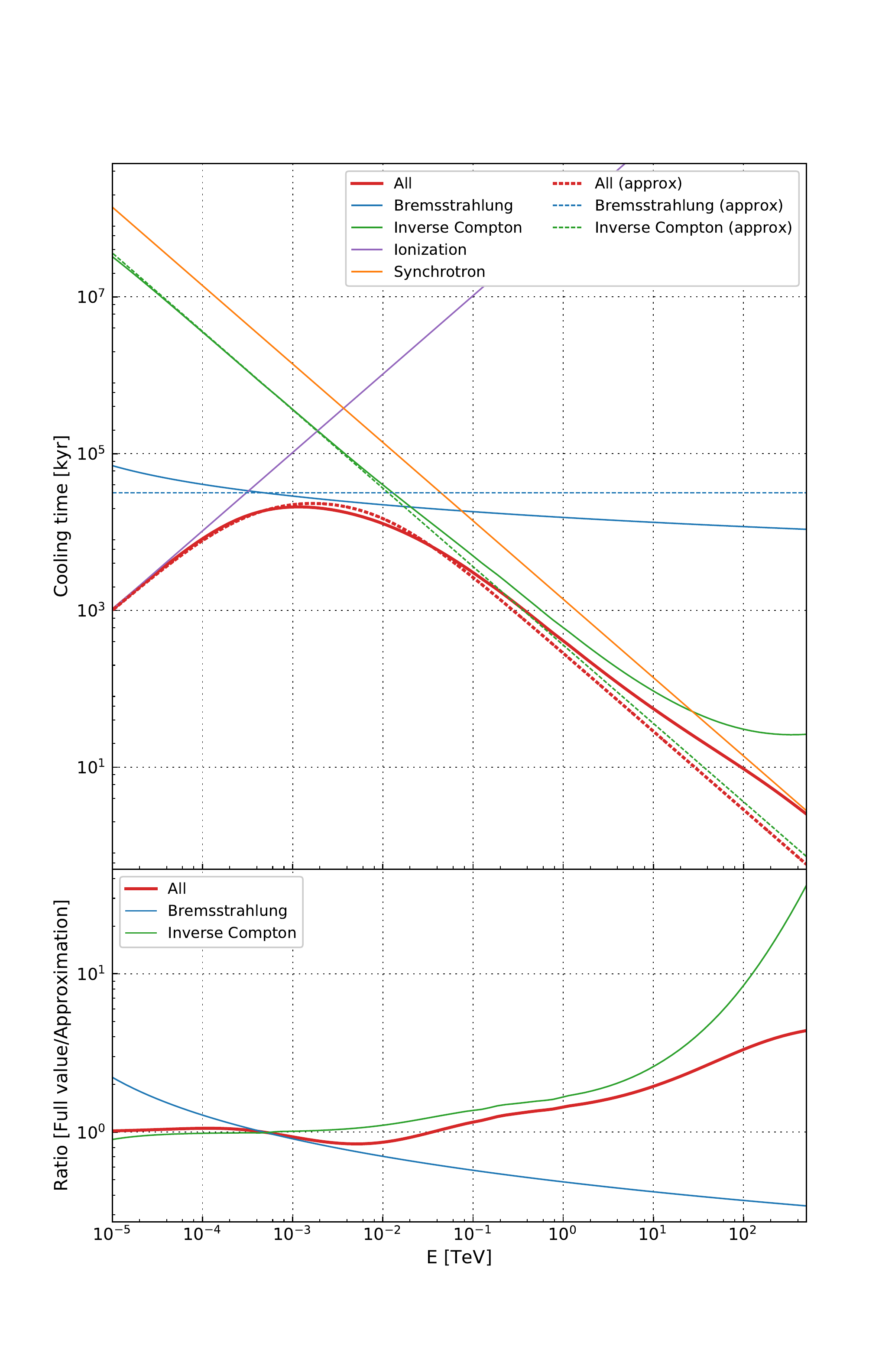}
\vspace*{-15mm}
\caption{Energy losses for different electron cooling mechanisms (top panel) using the classical Thomson approximation and taking into account the full KN formulation. In the bottom panel we have the ratio between the Thomson approximation and the KN value.}
\label{fig:losses_all}
\end{center}
\end{figure}

\begin{figure}
\begin{center}
\includegraphics[width=0.48\textwidth]{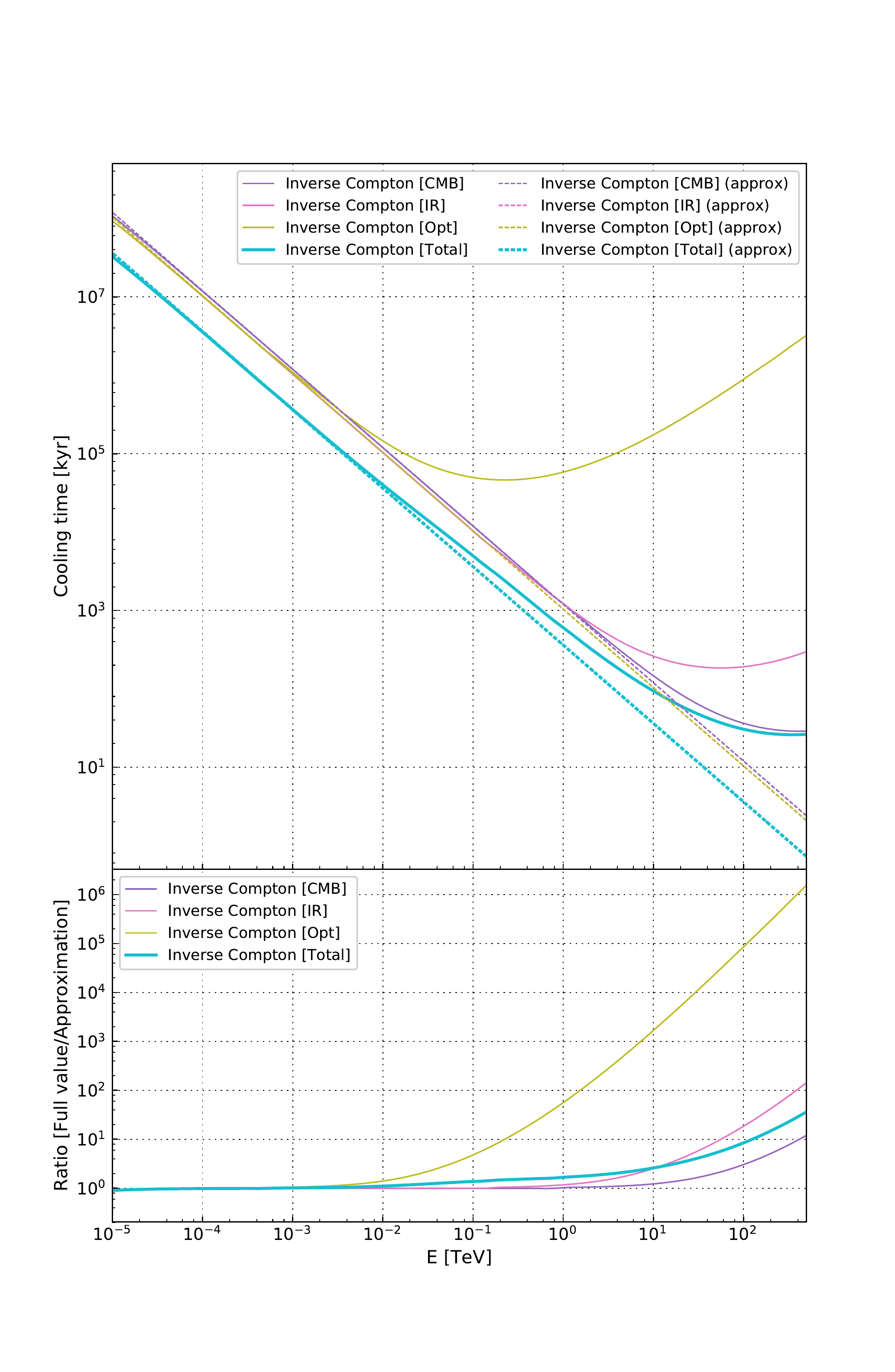}
\vspace*{-15mm}
\caption{IC energy losses for different photon fields (top panel) using the classical Thomson approximation and taking into account the full KN formulation. In the bottom panel we have the ratio between the Thomson approximation and the KN value.}
\label{fig:losses_ic}
\end{center}
\end{figure}

The energy losses for different cooling mechanisms are shown on Figure \ref{fig:losses_all}. For energies between  $E=10$ MeV and $\sim$1 GeV, where ionization and bremsstrahlung losses dominate, the electron cooling time is slightly overestimated using the aforementioned approximation. For energies between 1 GeV and 10 GeV the cooling is underestimated by a maximum value of 10\%. At energies above 10 GeV, the IC cooling starts to dominate over bremsstrahlung and the Thomson approximation gives an IC cooling ratio $<$2 for $E<10$ TeV. Since the cooling process that is more affected by using the classical value is the IC, we show them separately for different target photon fields on Figure \ref{fig:losses_ic}. We can see that the IC losses are already affected by KN losses for $E\sim$10 GeV due to the IC on optical photons. At $E\sim$1 TeV, the KN effects on IR photons start to deviate the cooling time from the Thomson value and for CMB photons this transition occurs at $E\sim10$ TeV. The energy losses due to synchrotron is equal because no approximation is used.

\begin{figure}
\begin{center}
\includegraphics[width=0.48\textwidth]{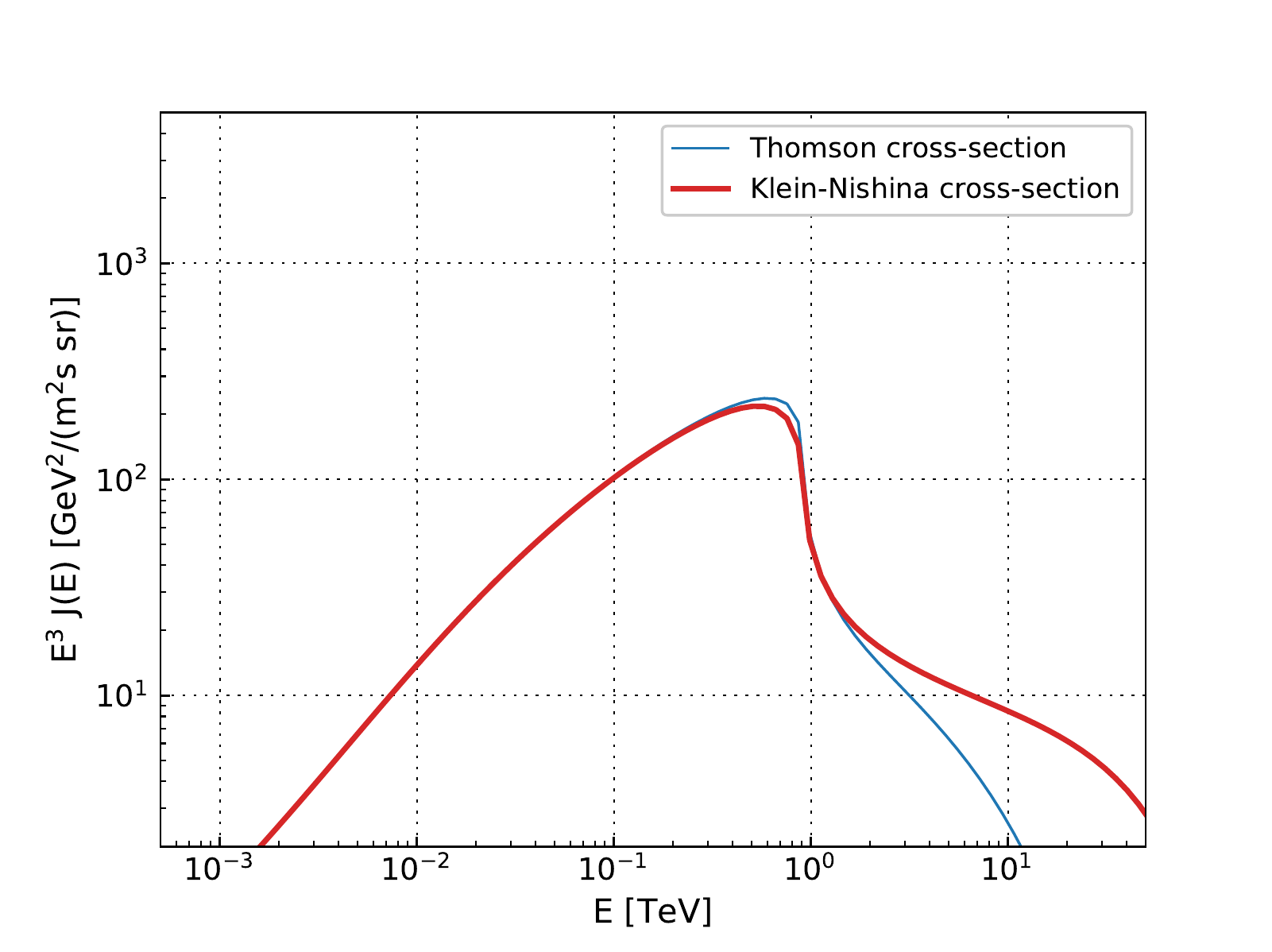}
\includegraphics[width=0.48\textwidth]{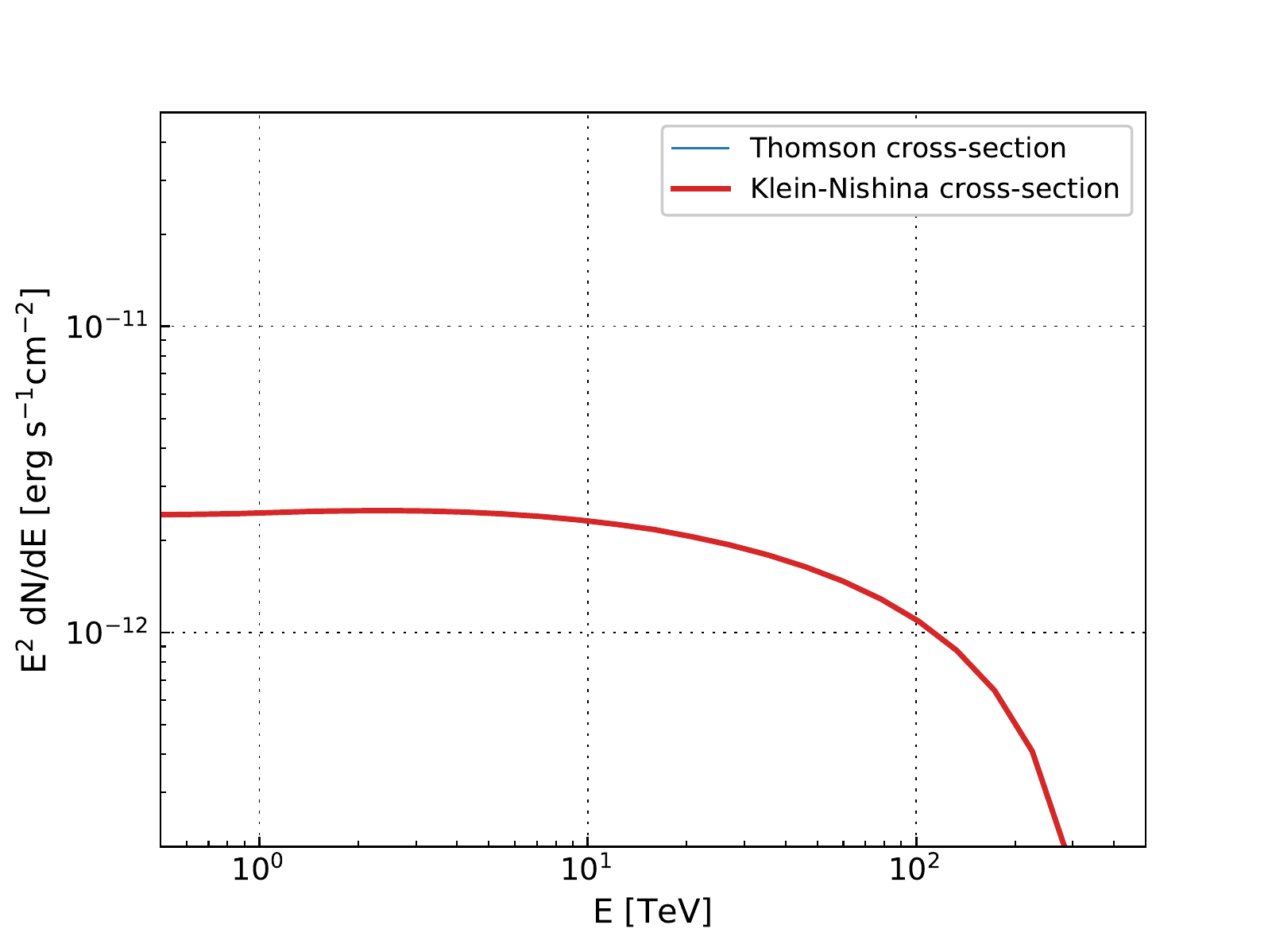}
\caption{All-electron flux at the Earth (top panel) and $\gamma$-ray spectrum (bottom panel) using the Thomson cross-section approximation and the full KN one.}
\label{fig:KN_effects_gamma_electrons}
\end{center}
\end{figure}

These deviations also affect the $\gamma$-ray spectrum and the e$^\pm$ at the Earth produced by the source. In particular, ionization and bremsstrahlung losses do not affect the energies we are discussing in this paper, but the KN effects are non-negligible for the absolute values derived. A comparison of the $\gamma$-ray spectrum and the e$^\pm$ at the Earth produced by a source using the full KN equation and the Thomson approximation are shown on Figure \ref{fig:KN_effects_gamma_electrons}. The e$^\pm$ flux at the Earth is not affected due to KN losses for $E<1$ TeV, but the flux detected at the Earth is larger when taking into account KN effects, producing a difference of more than one order of magnitude for $E>10$ TeV. This is explained by the suppression of the energy losses at these energies when using the KN formulation instead of the Thomson approximation. Since more electrons of high energy are present, the $\gamma$-ray flux is also higher in the case of using the full formula and the difference can be up to 50\% for energies between 10 and 100 TeV.

\subsection{Effect of different parameters on the $\gamma$-ray spectrum and all-electron flux}

We will evaluate what is the effect in the  all-electron flux at the Earth and the gamma-ray spectrum of varying several parameters in the diffusion code. In Figures \ref{fig:flux_earth_all} and \ref{fig:gamma_spectra_all} we have a comparison of what is the variation in the local all-electron flux and the $\gamma$-ray spectrum of a source when fixing all parameters except one. The red line, corresponding to the default values shown on Figure \ref{tab:diffusionparams}, remains unchanged in all the plots.

\begin{table}[ht]
  \begin{center}
  \begin{tabular}{|c|c|}
    \hline
    Parameter& Default value\\
     \hline
      Characteristic Age $\tau_{\rm{c}}$& 342 kyr \\
     \hline
      Distance $d_{\mathrm{Earth}}$& 250 pc \\
     \hline
      Dffusion exponent $\delta$ &  0.33 \\
      \hline
      Dffusion coefficient $D_0$ &  4$\times10^{27}$ cm$^{-2}$s$^{-1}$\\
     \hline
      Injection spectrum index $\alpha$ & 2.2 \\
     \hline
      Magnetic field $B$ &  3 $\mu$G\\
      \hline
      Maximum Energy $E_{\rm{max}}$ &  500 TeV \\
     \hline
      Minimum Energy $E_{\rm{min}}$ &  1 GeV \\
      \hline
      Spin-down fraction $\mu$  & 0.5 \\
      \hline        
            Breaking index $n$  & 3 \\
      \hline     
            Initial spin-down timescale $\tau_0$  & 10$^4$ yr \\
      \hline                          
  \end{tabular}
  \caption{Default values for the diffusion parameters.}
  \label{tab:diffusionparams}
  \end{center}
\end{table}

\subsubsection{Characteristic Age ($t_{\rm{age}}$)}
The age of the pulsar is not affecting the VHE $\gamma$-ray emission of the source. The reason is that the electrons producing this emission are already cooled and the system is in equilibrium for the range of ages considered. 
The emission at lower energies is affected by the age of the system since we enter into the energy range where we have cooled and uncooled electrons. 
The peak of the all-electron flux at the Earth is shifted to lower energies with increasing age. This peak separates the region where the diffusion is age-dominated ($t_{\rm{age}} < t_{\rm{cool}}(E)$, see Eq. \ref{eq:cooling}) to the one that is cooling dominated ($t_{\rm{age}} > t_{\rm{cool}}(E)$). We will denominate $E_{\rm{cool}}$ to the energy that fulfills $t_{\rm{age}} = t_{\rm{cool}}(E_{\rm{cool}}$)
The total all-electron flux also slightly increases with increasing age due to the fact that electrons have diffused further away after larger times.

\subsubsection{Distance ($d_{\rm{Earth}}$)}

For a given energy $\gamma$ and characteristic age $t_{\rm{age}}$, the energy density of electrons at the Earth for a distance $d$ is proportional to:

\begin{equation}
f(d,t_{\rm{age}},\gamma) \propto \exp \left( - \frac{d^2}{r_{\textrm{diff}}^2} \right)
\end{equation}

where $r_{\textrm{diff}}(\gamma,t_{\rm{age}})$ is the diffusion radius for an energy $\gamma$. The ratio between the energy density of electrons at two different distances $d_{\rm{a}}$ and $d_{\rm{b}}$ is given by:

\begin{equation}
\frac{f(d_{\rm{b}},t_{\rm{age}},\gamma)}{f(d_{\rm{a}},t_{\rm{age}},\gamma)} = \frac{\exp \left( - \frac{d_{\rm{b}}^2}{r_{\textrm{diff}}^2} \right)}{\exp \left( - \frac{d_{\rm{a}}^2}{r_{\textrm{diff}}^2} \right)} = \exp \left(  \frac{d_{\rm{a}}^2 - d_{\rm{b}}^2}{r_{\textrm{diff}}^2}  \right)
\end{equation}

The feature that appears at $E>E_{\rm{cool}}$ comes from the injection of electrons in a dipole form (see \cite{Atoyan95,Aharonian95,Yuksel08} for similar results for $E>E_{\rm{cool}}$).

The difference in VHE $\gamma$-ray emission comes from two effects: if all the electrons emitted by the same source are integrated, its $\gamma$-ray flux when it is located at a distance $d_{\rm{a}}$, scales with that of the same source situated at a distance $d_{\rm{b}}$ as:

\begin{equation}
\frac{f(\gamma,d_{\rm{a}})}{f(\gamma,d_{\rm{b}})} \propto \left( \frac{d_{\rm{a}}}{d_{\rm{b}}}\right)^2
\end{equation}

Since the size of the sources is finite, we would integrate more electrons from sources located at larger distances, so the difference in the VHE $\gamma$-ray spectra is between 1 and this ratio.

\subsubsection{Dffusion coefficient normalization $D_0$}
\label{sec:d0}

According to Eq. \ref{eq:energy_density}, the energy density is proportional to $f(x)=x^3 \exp(-x^2)$, where $x=d_{\rm{Earth}}/r_{\textrm{diff}}$. $f(x)$ is a function with a maximum at $x=\sqrt{3/2}$, meaning that for a given distance, the energy density increases with the diffusion radius until $r_{\textrm{diff}} = \sqrt{2/3} d_{\rm{Earth}}$ where it drops abruptly.
The diffusion radius is proportional to the diffusion coefficient, given by Eq. \ref{eq:diff_coefficient}, which increases with increasing $D_0$. 
For a given distance and energy, the energy density will increase with increasing $D_0$ until the diffusion radius reaches its limit $\sqrt{2/3} d$. For larger diffusion radii, the energy density decreases with respect to that obtained for lower diffusion coefficients.
This effect can be seen in Figure \ref{fig:flux_earth_all} where the energy density of electrons increases with $D_0$, to start decreasing for energies where $r_{\textrm{diff}} \approx \sqrt{2/3} d$.

In the region for $E>E_{\rm{cool}}$, the system is cooling dominated, the diffusion radius decreases and the succession of energy densities comes back to the order of higher energy density for higher diffusion coefficients.





Regarding the VHE $\gamma$-ray spectrum, the difference between different curves comes from the finite size of the source and the different speed at which electrons are diffusing in each of the cases. If we increase the integration radius or decrease the diffusion coefficient in a way that all the electrons are included within the line of sight integral, the VHE $\gamma$-ray emission should be the same for all the cases. The trend in $D_0$ is explained as follows: for the same source size, the faster the diffusion, the small quantity of electrons producing gamma rays inside the line of sight.

\subsubsection{Diffusion index $\delta$}
 
As in the previous section, the diffusion coefficient and diffusion radius increase with increasing $\delta$. The energy density is slightly larger for larger $\delta$ for $E$ fulfilling $r_{\textrm{diff}}<< d_{\rm{Earth}}$. With increasing $E$, when $r_{\textrm{diff}}$ approaches $d_{\rm{Earth}}$, the energy density function changes its shape and increases at a lower rate. If the diffusion radius is large enough (producing a decrease in the energy density when $r_{\textrm{diff}}>2/3 d_{\rm{Earth}}$) to overcome the increase produced by the increasing energy, the maximum of the energy density is shifted to energies lower than $E_{\rm{cool}}$.

For the VHE $\gamma$-ray emission we see the same effects as in the previous section. When the diffusion coefficient increases for a given energy, particles diffuse further away and therefore are not computed in the line of sight integral, making the VHE $\gamma$-ray flux of the source lower.

\subsubsection{Injection spectrum index $\alpha$}

When we vary $\alpha$ we see a similar effect as shown in Section \ref{sec:d0}. In that case, for a given energy, the variation was produced in the ratio $x=d_{\rm{Earth}}/r_{\textrm{diff}}$ due to a variation in the diffusion coefficient. Here we change the energy-dependence of the injection spectrum. For larger values of $\alpha$, the electron spectrum decreases faster with energy and the energy quantity injected at higher energies is lower. The flux at the Earth therefore softens with increasing energy faster for larger $\alpha$. This effect is independent of the cooling of the electrons, being more pronounced at higher energies.


Since less electrons are injected at higher energies, the VHE $\gamma$-ray spectrum follows the same trend as the electron spectrum.

\subsubsection{Magnetic field $B$}
The magnetic field affects the cooling time of the electrons. If we consider the cooling time approximation given by Eq. \ref{eq:cooling}, for larger magnetic fields the cooling time will be lower for a given energy. The maximum of the energy density, located at $E_{\rm{cool}}$ moves to lower energies for increasing $B$, since $E_{\rm{cool}}$ is also reduced. The total energy density is also reduced since electron losses are larger.

The variation in the magnetic energy density is given by:

\begin{equation}
U_{\rm{mag}} = \frac{B^2}{8\pi}=0.22\left(  \frac{B}{3 \mu\rm{G}} \right) \rm{eV cm}^{-3}
\end{equation}

therefore, the energy at which the age equals the cooling time will be given by:

\begin{equation}
E_{\rm{cool}} = \frac{3\times10^5}{t_{\rm{age}}} \frac{U_{\rm{mag}} + U_{\rm{ph}}}{\rm{eV cm}^{-3}}
\end{equation}

where $U_{\rm{ph}}\approx1 \rm{eV cm}^{-3}$

Since there are less electrons at high energies for larger $B$ due to a higher cooling, the VHE $\gamma$-ray emission at these energies is also lower.

\subsubsection{Braking index $n$}
\label{sec:braking_index}
If the pulsar loses its energy as a dipole, the braking index $n$ should be 3. There are very few braking indices measured for pulsars and they are all smaller than the pure dipole value: in the range between 1.41--2.91. 
For older pulsars, the braking index might be larger than this value. 
We tested different braking indices as it is shown in Figures \ref{fig:flux_earth_all} and \ref{fig:gamma_spectra_all}. 


Higher energies are less affected by the variation of $n$ because they depend on more freshly injected electrons. Since the spin-down power of the pulsar at $t_{\rm{age}}$ is the same in every case and the spectrum of these electrons is cooling limited, the spin-down power of the pulsar at a time close to $t_{\rm{age}}$ is similar for all the braking indices tested. The electrons are injected at times closer to the age of the pulsar with increasing energies, therefore they are less affected by the difference on braking indices. Since a change in $n$ implies a change in $t_{\rm{age}}$, according to equation \ref{eq:age}, the shift of the peak on the flux at the Earth to higher energies for increasing $n$ is a result of that.
The $\gamma$-ray spectrum of the source above $E>1$ TeV is produced by electrons of $E>10$ TeV. Since the injection of these electrons was very recent compared to the age of the pulsar and $\tau_0$, the total quantity of electrons integrated within the area considered for $\gamma$-ray emission is similar in every case.

\subsubsection{Initial spin-down timescale $\tau_0$}

The initial spin-down timescale is a very uncertain quantity on the evolution of the pulsar as well. It can be understood as the time interval from the birth of the pulsar during which it releases $\sim$half of its total power. Initial spin-down timescales considered in the literature are in the range of 1-10 kyr. 


We can see that $\tau_0$ has a similar behaviour as $n$ in the all-electron flux at the Earth and $t_{\rm{age}}$: with increasing $\tau_0$, the all-electron flux at the Earth for cooled electron decreases. The VHE $\gamma$-ray spectrum remains unaffected for the same reason as in Section \ref{sec:braking_index}.

\subsubsection{Percentage of spin-down power that is transformed into electrons $\mu$}
The effect of varying the following parameters are not included on Figures \ref{fig:flux_earth_all} and \ref{fig:gamma_spectra_all}. The reason is that the effect on the $\gamma$-ray spectrum and the all-electron flux at the Earth is only a shift proportional to the quantities given. In the case of $\mu$, the effect on both the $\gamma$-ray and the all-electron spectrum is a multiplication factor by the $\mu$ factor used. This is also equivalent to vary the $\dot{E}$ of the pulsar.

\subsubsection{Minimum Energy $E_{\rm{min}}$}

The minimum and maximum energy of the electrons affect the normalization of the spectrum, since they determine the energy range where the total energy injected by the pulsar is distributed. For a injection spectrum of the form presented in Eq. \ref{eq:injection_spectrum}, the total energy injected $E_{\rm{g}}$ can be calculated as:

\begin{equation}
E_{\rm{g}}=\int^{\rm{Emax}}_{\rm{Emin}}{\frac{dN}{dE} E dE} = Q_0 (E_{\rm{max}}^{-\alpha+2}-E_{\rm{min}}^{-\alpha+2})
\end{equation}

for $\alpha \neq 2$. In the case we are studying, where the default $\alpha>2$ , and $E_{\rm{max}}>>E_{\rm{min}}$ the equation for the total energy might be simplified as:

\begin{equation}
\label{eq:total_energy_normalization}
E_{\rm{g}}=\frac{Q_0}{E_{\rm{min}}^{\alpha-2}} 
\end{equation}

Since $E_{\rm{g}}$  is constant, the normalization $Q_0$ is only dependent on $E_{\rm{min}}$ and $\alpha$. If we vary $E_{\rm{min}}$ from $E_{\rm{min,a}}$ to $E_{\rm{min,b}}$, the energy density at the Earth will be affected by the quantity $(E_{\rm{min},b}/E_{\rm{min,a}})^{(\alpha-2)}$. The VHE $\gamma$-ray emission is affected in the same way as the energy density of the electrons at the Earth.

\subsubsection{Maximum Energy $E_{\rm{max}}$}

According to Eq. \ref{eq:total_energy_normalization}, the maximum energy of the electrons does not affect the normalization of the spectrum. The normalization is not affected for $E<5$ TeV, but at higher energies, there is a slight increase for higher $E_{\rm{max}}$ coming simply from the extension of the injection spectrum to higher energies. The same effect is reproduced in the VHE $\gamma$-ray spectrum.

\begin{figure*}
\begin{center}
\vspace*{-50mm}
\includegraphics[width=1.1\textwidth]{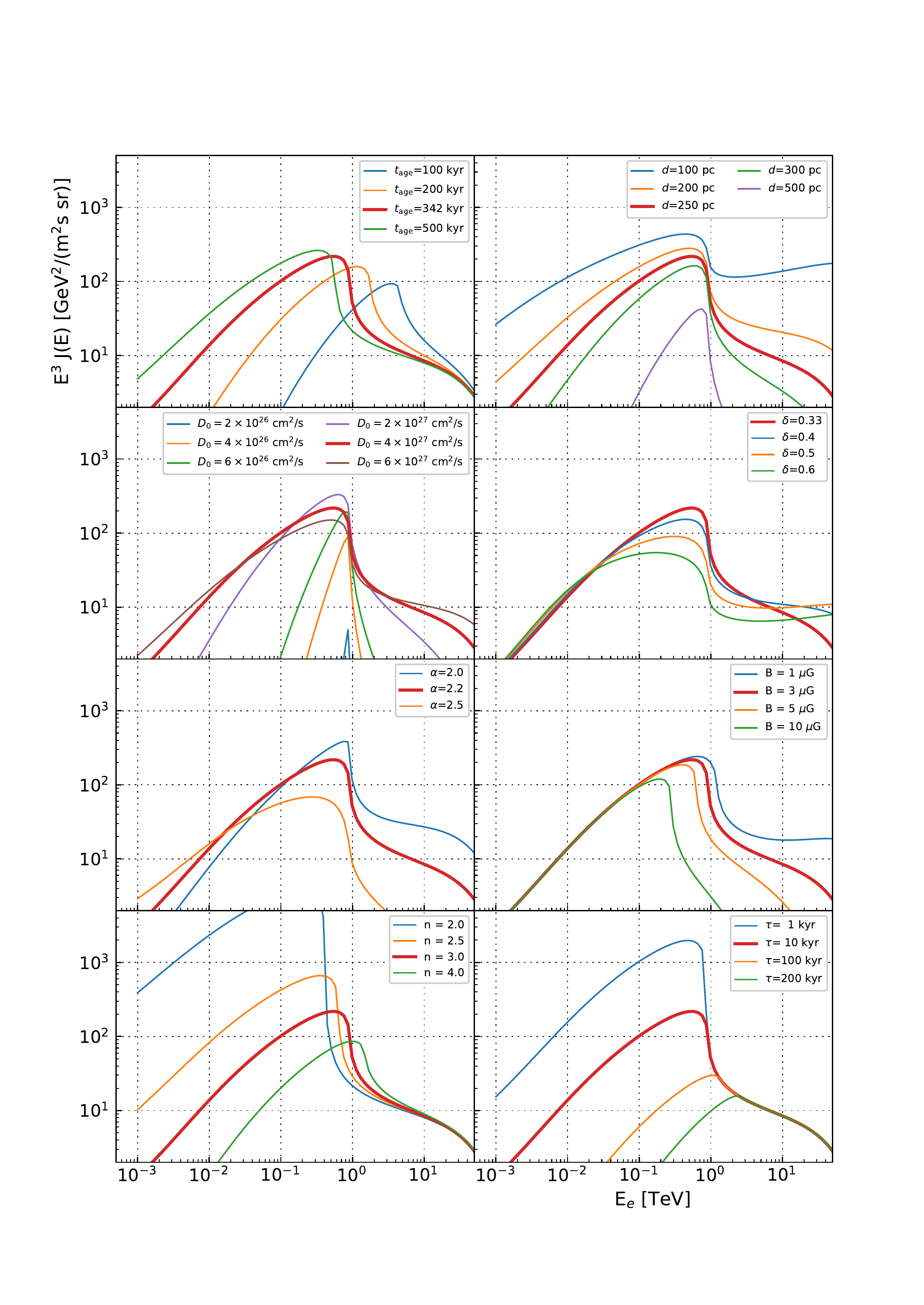}
\vspace*{-20mm}
\caption{Electron/positron flux at the Earth from a central source injecting electrons using the default values from Table \ref{tab:diffusionparams}, with the variations indicated in each of the legends. The red lines correspond to the default value used throughout the paper.}
\label{fig:flux_earth_all}
\end{center}
\end{figure*}

\begin{figure*}
\begin{center}
\vspace*{-50mm}
\includegraphics[width=1.1\textwidth]{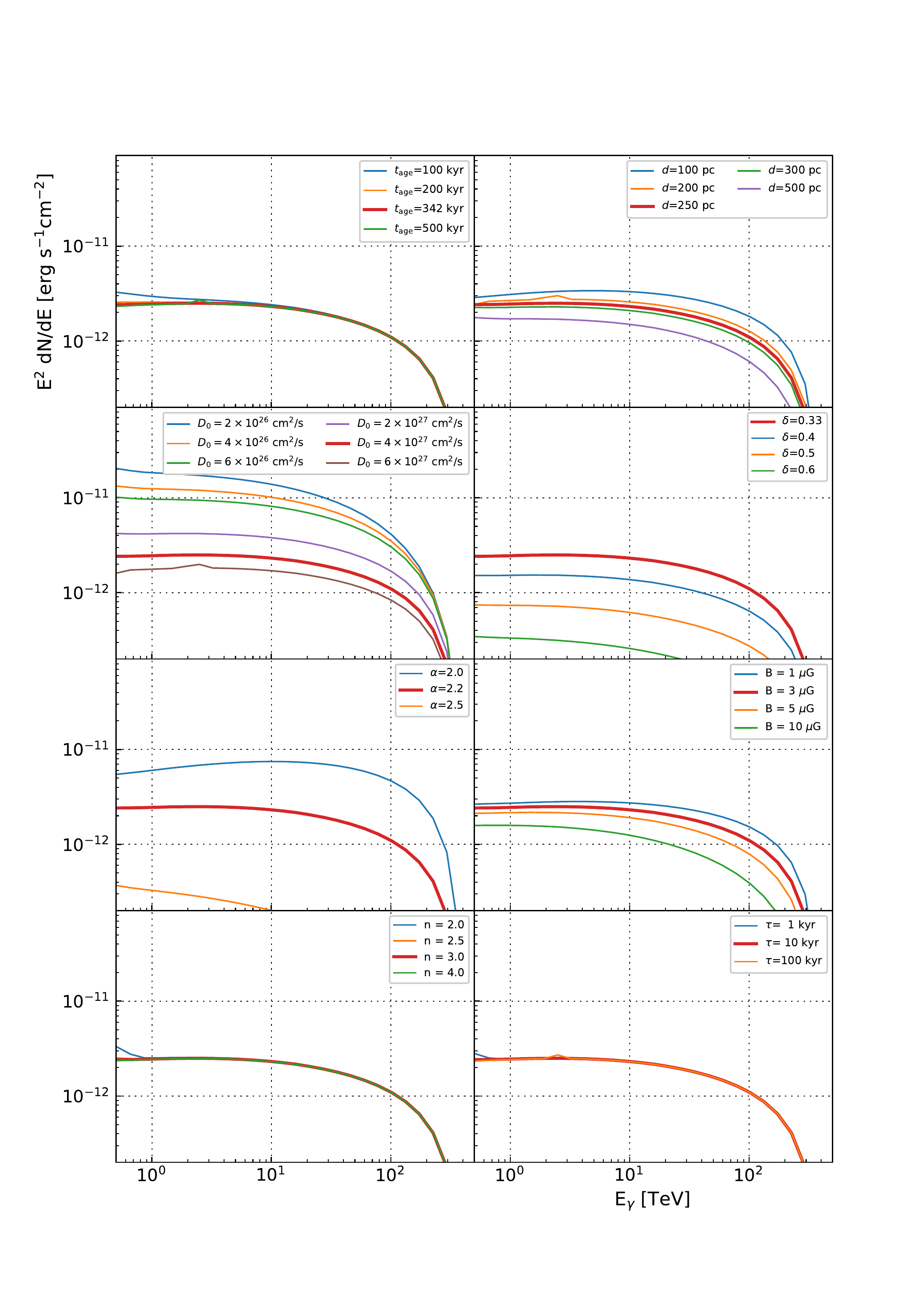}
\vspace*{-20mm}
\caption{Gamma spectra from a central source injecting electrons using the default values from Table \ref{tab:diffusionparams}, with the variations indicated in each of the legends.  The red lines correspond to the default value used throughout the paper.}
\label{fig:gamma_spectra_all}
\end{center}
\end{figure*}

\subsection{Effect of the birth period}

\subsubsection{Fixing birth period}
We studied the effect of changing the birth period $P_0$ while keeping the total energy injected by the pulsar constant. In the bottom panels of Figures \ref{fig:flux_earth_all} and \ref{fig:gamma_spectra_all} we show the difference on the all-electron flux at the Earth and the VHE gamma-ray spectrum of a source when varying the braking index $n$ and the initial spin-down timescale $\tau_0$. While showing these differences is illustrative for the effects on the particle diffusion, the total energy injected by the central source is not constant. The approach we take in this section is, given a central source that has injected a fixed amount of energy during its lifetime, what is the difference in the all-electron flux at the Earth and VHE gamma-ray spectrum when varying these parameters. According to 6 and 7 from \cite{Gaensler06}, the initial birth period of a pulsar $P_0$ is related to its current period ($P$), characteristic age ($\tau_{\mathrm{c}}$) and initial spin-down timescale ($\tau_0$) by:

\begin{equation}
\label{eq:P_P0}
\left(\frac{P}{P_0}\right)^{n-1} = \frac{2 \tau_{\mathrm{c}}}{\tau_0(n-1)}
\end{equation}

\begin{figure}
\begin{center}
\includegraphics[width=0.48\textwidth]{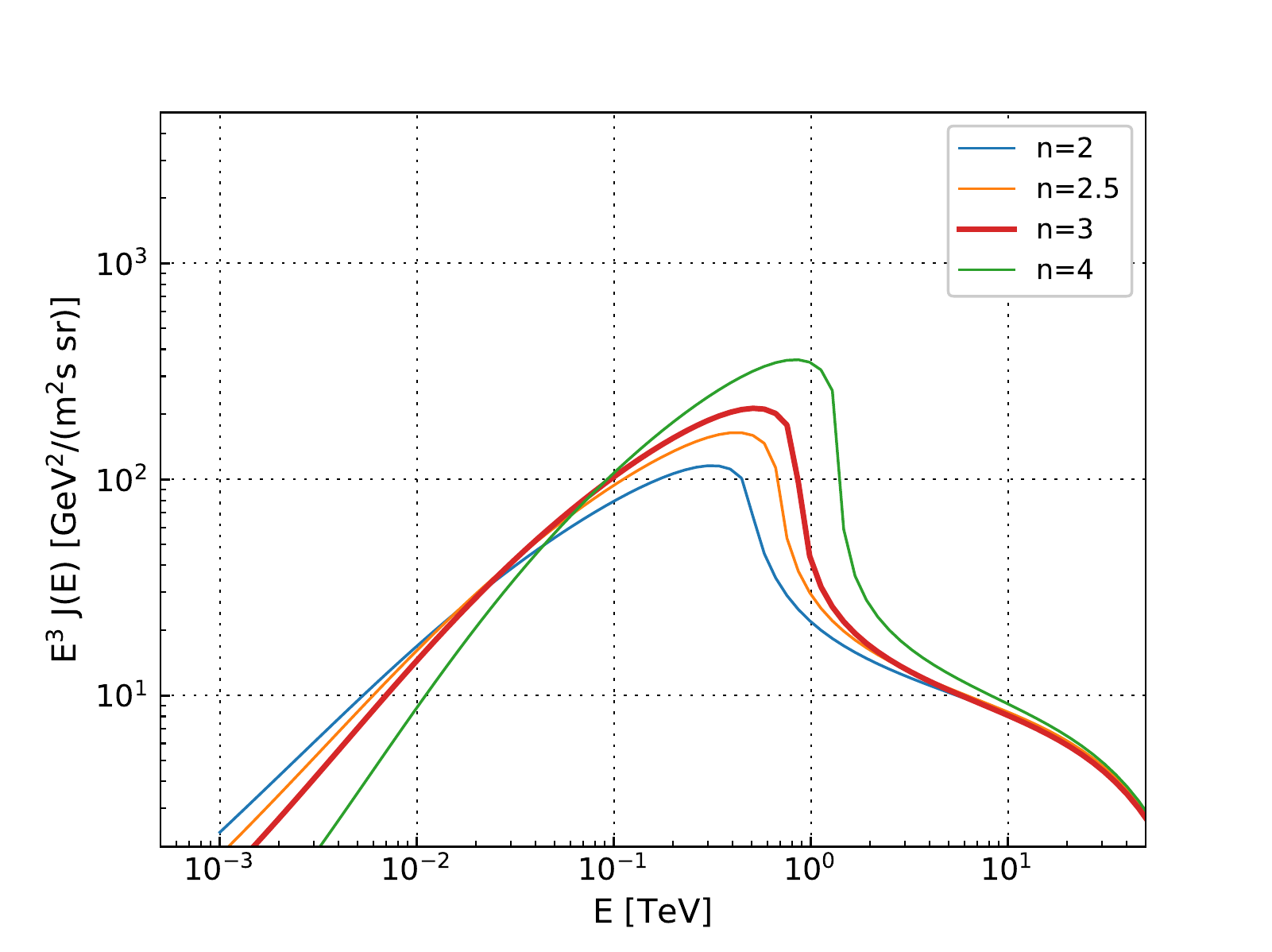}
\caption{Electron/positron flux at the Earth for different braking indices keeping constant $P_0, P, \dot{E}$ and $\tau_{\rm{c}}$.}
\label{fig:flux_Earth_brind}
\end{center}
\end{figure}

If we take the pulsar of our example that has $P$=237 ms, the birth period for the parameters shown in Table \ref{tab:diffusionparams} is $P_0=40.5$ ms. If we fix $P_0$ and $P$, the total energy injected by the pulsar will be fixed. If we additionally fix $\tau_{\rm{c}}$ and $\dot{E}$, the variation in $n$ will show how this energy was injected, without varying the total quantity. In Figure \ref{fig:flux_Earth_brind} we can see the variation of the all-electron flux at the Earth for different braking indices. Modifying $n$ by keeping all the other parameters constant, implies a modification of $\tau_0$, that is smaller for larger $n$. Larger $n$ therefore implies an injection that is closer to the burst-like approximation and a larger flux at the Earth for $E<E_{\rm{cool}}$. The variation in $n$ is also affecting $t_{\rm{age}}$, according to equation \ref{eq:age}. The age decreases with increasing $n$, hence the shift to higher energies of the peak for larger $n$.

Note that these results are not against to what is shown in the bottom panels of Figure \ref{fig:flux_earth_all}. There, we evaluated the effect of a variation in the braking index $n$ letting free the total energy injected by the pulsar. This implies a larger initial luminosity $L_0$ according to equation \ref{eq:luminosity}, which added to the same evolution parameters results into more energy injected by the pulsar. Just as an example of the amount of energy injected in these cases, for $n=2$ shown in Figure \ref{fig:flux_earth_all}, the initial birth period would be 3.5 ms, more than one order of magnitude lower than the one for $n=3$. 

\subsubsection{Varying birth period}

\begin{figure}
\begin{center}
\includegraphics[width=0.48\textwidth]{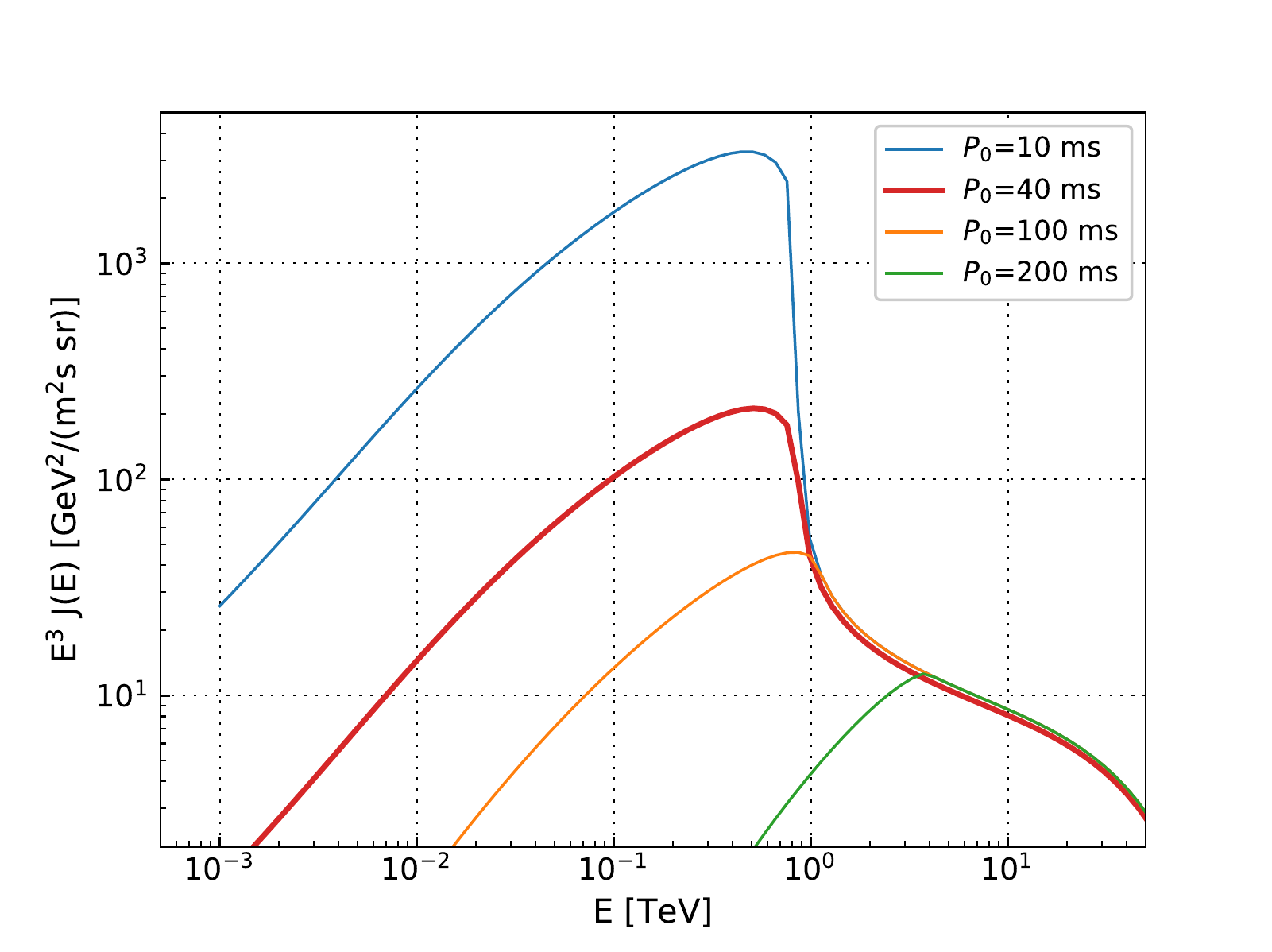}
\caption{Electron/positron flux at the Earth for different initial birth periods keeping constant $n, P, \dot{E}$ and $\tau_{\rm{c}}$.}
\label{fig:flux_Earth_P0}
\end{center}
\end{figure}

Similarly to the previous section, if we keep constant the braking index, $P$, $\tau_{\rm{c}}$ and $\dot{E}$ and vary the birth period of the pulsar ($P_0$), we will be modifying the total energy that is injected into the system. Since the total energy of a rotator is given by:

\begin{equation}
E=\frac{1}{2} I \omega^2 
\end{equation}

where $\omega=\frac{2\pi}{P}$, the total energy injected by the pulsar, when keeping $P$ constant, is proportional to 1/$P_0^2$. Keeping all the aforementioned values constant and varying $P_0$ implies a variation in $\tau_0$, smaller for smaller $P_0$, following equation \ref{eq:P_P0}. This also implies an injection closer to burst-like for lower $P_0$, and a $t_{\rm{age}}$ lower for lower $P_0$ as well, as it is shown in Figure \ref{fig:flux_Earth_P0}. 

\section{Conclusion}
We developed a code to calculate the diffusion of electrons and positrons from point-like sources to the Earth. With the code we can calculate the distribution of electrons and positrons produced by a central source, the $\gamma$-ray spectrum produced by this source and the electron and positron flux produced by these source at the Earth, as well as the positron fraction. We studied the effects of particle suppression, cooling approximations and the variation of different pulsar and diffusion parameters on the local all-electron flux and the VHE $\gamma$-ray spectrum produced by a source.

The code can be found in the github repository:
\\
https://github.com/rlopezcoto/EDGE

\section*{Acknowledgements}
The authors would like to thank the HAWC collaboration for useful discussions during the development of the electron diffusion code.

\section*{References}
\bibliographystyle{./style/elsarticle/elsarticle-num_etal} 

 \bibliography{./references}
\end{document}